\documentclass{amsart}
\usepackage{amsbsy,amssymb,amscd,amsfonts,latexsym,amstext,delarray,
amsmath,amsthm,graphicx}

\input xypic

\numberwithin{equation}{section}

\def\bG{{\mathbb G}}

\def\bK{{\mathbb K}}

\def\bT{{\mathbb T}}

\def\A{{\mathbb A}}
\def\C{{\mathbb C}}
\def\F{{\mathbb F}}
\renewcommand{\H}{{\mathbb H}}
\def\N{{\mathbb N}}
\renewcommand{\P}{{\mathbb P}}
\def\Q{{\mathbb Q}}
\def\Z{{\mathbb Z}}
\def\R{{\mathbb R}}
\def\K{{\mathbb K}}

\def\fb{{\mathfrak b}}
\def\fn{{\mathfrak n}}
\def\fa{{\mathfrak a}}
\def\ff{{\mathfrak f}}

\def\cA{{\mathcal A}}

\def\cD{{\mathcal D}}

\def\cF{{\mathcal F}}

\def\cH{{\mathcal H}}

\def\cL{{\mathcal L}}
\def\cM{{\mathcal M}}

\def\cP{{\mathcal P}}

\def\cS{{\mathcal S}}
\def\cT{{\mathcal T}}

\newcommand{\cf}{{\it cf.\/}\ }
\def\text{\hbox}

\def\Aut{{\rm Aut}}

\def\End{{\rm End}}

\def\Gal{{\rm Gal}}
\def\GL{{\rm GL}}

\def\PSL{{\rm PSL}}

\def\SL{{\rm SL}}

\def\Tr{{\rm Tr}}

\title{Noncommutative geometry and arithmetic}
\author{Matilde Marcolli}
\address{Division of Physics, Mathematics, and Astronomy  \\
California Institute of Technology \\ Mail Code 253-37, 1200 E.California Blvd \\
Pasadena, CA 91125, USA}
\email{matilde\@@caltech.edu}

\begin{document}
\maketitle

\begin{abstract}
This is an overview of recent results aimed at developing a geometry of
noncommutative tori with real multiplication, with the purpose of providing
a parallel, for real quadratic fields, of the classical theory of elliptic curves 
with complex multiplication for imaginary quadratic fields. This talk
concentrates on two main aspects: the relation of Stark numbers to
the geometry of noncommutative tori with real multiplication, and the
shadows of modular forms on the noncommutative boundary of modular 
curves, that is, the moduli space of noncommutative tori. 
\end{abstract}

\section{Introduction}

The last few years have seen the development of a new 
line of investigation, aimed at applying methods of noncommutative
geometry and theoretical physics to address questions in number theory. A broad 
overview of some of the main directions in which this
area has progressed can be found in the recent monographs
\cite{Mar-book} and \cite{CoMa-book}.  In this talk I am going to
focus mostly on a particular, but in my opinion especially promising,
aspect of this new and rapidly growing field, which did not get
sufficient attention in \cite{CoMa-book}, \cite{Mar-book}: the question of developing
an appropriate geometry underlying the abelian extensions of real
quadratic fields. This line of investigation was initially proposed by
Manin in \cite{Man}, \cite{Man2}, as the ``real multiplication program"
and it aims at developing within noncommutative geometry
a parallel to the classical theory of
elliptic curves with complex multiplication, and their role in the
explicit construction of abelian extensions of imaginary quadratic
fields, which would work for real quadratic fields. 
I am going to give an overview of the current state of the
art in addressing this problem, by focusing on those aspects I
have been more closely involved with.

\smallskip

There are two complementary approaches to developing a noncommutative
geometry of real quadratic fields. One is based on working with
noncommutative tori as substitutes for elliptic curves, focussing on 
those whose real parameter is a quadratic irrationality, which have
non trivial self Morita equivalences, analogous to the complex
multiplication phenomenon for elliptic curves. This approach requires
constructing suitable functions on these spaces, which replace the
coordinates of the torsion points of elliptic curves, hence the problem
of finding suitable algebraic models for noncommutative tori. 
I will concentrate here especially on the question of how to express certain
numbers, the Stark numbers, which conjecturally generate abelian
extensions of real quadratic fields, in terms of the geometry of
noncommutative tori.

\smallskip

The other complementary approach deals with a noncommutative
space that parameterizes noncommutative tori up to Morita equivalence.
This is sometimes referred to as the ``invisible boundary" of the modular
curves, since it parameterizes those degenerations of elliptic
curves with level structure that are no longer expressible in algebro-geometric
terms but that continue to exist as noncommutative tori. A related adelic
version includes degenerations of the level structure and gives rise
to a quantum statistical mechanical system based on the commensurability
relation of lattices with possibly degenerate level structures, whose
zero temperature equilibrium states, evaluated on an algebra
of arithmetic elements should conjecturally provide generators 
of abelian extensions. The main problem in this approach is to obtain
the right algebra of functions on this invisible boundary, which should
consist of holographic images, or ``shadows", that modular forms
on the bulk space cast upon the invisible boundary.

\section{Elliptic curves and noncommutative tori}

Elliptic curves are among the most widely studied objects in
mathematics, whose pervasive presence in geometry,
arithmetic and physics has made them a topic of nearly
universal interest across mathematical disciplines. In
number theory, one of the most famous manifestations
of elliptic curves is through the theory of complex
multiplication and the abelian class field theory
problem (Hilbert 12th problem) in the case of
imaginary quadratic fields.

\smallskip

The analytic model of an elliptic curve is the complex manifold
realized as a quotient
$E_\tau (\C)= \C^2 /\Lambda$ with $\Lambda = \Z + \Z \tau$ or with the
Jacobi uniformization $E_q(\C)=\C^* /q^\Z$ with $|q|<1$.
The endomorphism ring of an elliptic curves is a copy of $\Z$, except in the
special case of elliptic curves with complex multiplication where
$\End(E_\tau)=\Z + f O_\K$, with $O_\K$ the ring of integers of an
imaginary quadratic field and $f\geq 1$ an integer (the conductor).

\medskip

A beautiful result in number theory relates the geometry of
elliptic curves with complex multiplication to the explicit
class field theory problem for imaginary quadratic fields:
the explicit construction of generators of abelian extensions
with the Galois action. 

\smallskip

There are two formulations of this construction, one that
works directly with the CM elliptic curves, and the coordinates
of their torsion points, and one that works with the values
of modular forms on the CM points in the moduli space
of elliptic curves. (We refer the reader to \cite{Lang},
\cite{Shimura} for more information on this topic.)

\smallskip

As I will explain in the rest of the paper, both approaches 
have a noncommutative geometry analog in the case of
real quadratic fields, which is in the process of being
developed into a tool suitable for the investigation of the
corresponding class field theory problem.

\smallskip

In the elliptic curve point of view, one knows that the
maximal abelian extension $\K^{ab}$ of an imaginary
quadratic field $\K=\Q(\sqrt{-d})$ has explicit generators
$$ \K^{ab} = \K (t(E_{\K,{\rm tors}}), j(E_{\K})), $$
where $t$ is a coordinate on the quotient
$E_\K /\Aut(E_\K)\simeq \P^1$ and $j(E_\K)$ is the
$j$-invariant. 

\smallskip

I will explain below, based on a result of \cite{Mar},
how one can obtain an analog of the quotient
$E_\K /\Aut(E_\K)$ in the noncommutative geometry
context for real quadratic fields. I will also mention some
current approaches aimed at identifying the correct 
analog of the $j$-invariant in that setting.

\smallskip

Currently, the main problem in extending this
approach to real quadratic fields via noncommutative
geometry lies in the fact that, while elliptic curves
have, besides the analytic model as quotients, 
an algebraic model as algebraic curves defined
by polynomial equations, their noncommutative
geometry analogs, the noncommutative tori, have
a good analytic model, but not yet a fully
satisfactory algebraic model. I will comment more
on the current state of the art on this question in
\S \ref{analalg} below.

\smallskip

The other point of view, based on the moduli space,
considers all elliptic curves, parameterized by the
modular curve $X_\Gamma(\C)= \H/\Gamma$,
with $\H$ the complex upper half plane and
$\Gamma=\SL_2(\Z)$ acting on it by fractional linear
transformations. One considers then the field $F$ of
modular functions. In this setting, the explicit
class field theory result for imaginary quadratic
field is stated in terms of the generators
$$ \K^{ab}  = \K( f(\tau), \,\, f\in F, \,\,\, \tau\in \text{ CM points of } X_\Gamma ), $$
the values of modular functions at CM points. The Galois action
of $\Gal(\K^{ab}/\K)$ is induced by the action of the automorphism
group $\Aut(F)$ of the modular field. 

\medskip

The case of the explicit class field theory of $\Q$, the Kronecker--Weber
theorem, can be formulated in terms of a special degenerate case of
elliptic curves. When the parameter $q$ in the elliptic curve $E_q(\C)$ tends to a root
of unity, or equivalently when the parameter $\tau \in \H$ tends to a 
rational points in the real line, the elliptic curve degenerates to
a cylinder, the multiplicative group $\C^* =\bG_m(\C)$. The maximal
abelian extension of $\Q$ is then generated by the torsion points of
this degenerate elliptic curve,
$$ \Q^{ab} =\Q ( \bG_{m, {\rm tors}} ), $$
that is, by the roots of unity, the cyclotomic extensions. 

\medskip

The first case of number fields for which a 
solution to the explicit class field theory 
problem is not known is that of the real quadratic
fields $\K=\Q(\sqrt{d})$. The approach currently
being developed via noncommutative geometry
is based on the idea of relating this case also
to a special degenerate case of elliptic curves,
the {\em noncommutative tori}. 
Manin's ``Real multiplication program" \cite{Man}, \cite{Man2}, to which
I will return in the following, aims at building for
noncommutative tori a parallel to the theory of
complex multiplication for elliptic curves.

\medskip

When the modulus $q$ of the elliptic curve $E_q(\C)$ tends to a
point $\exp(2\pi i \theta)$ on the unit circle $S^1\subset \C^*$ which
is not a root of unity, or equivalently when $\tau\in \H$ tends to an
irrational point on the real line, the elliptic curve degenerates in
a much more drastic way. The action of $\Z$ by irrational rotations
on the unit circle has dense orbits, so that the quotient, in the usual
sense, does not deliver any interesting space that can be used
to the purpose of doing geometry. This prevents one from considering
such degenerations of elliptic curves in the usual algebro-geometric
or complex-analytic world. 

\medskip 

Noncommutative geometry, however, is explicitly designed in such
a way as to treat ``bad quotients" so that one can continue to make
sense of ordinary geometry on them as if they were smooth objects.
The main idea of how one does that is, instead of collapsing points
via the equivalence relation of the quotient operation, one keeps
all the identifications explicit in the groupoid describing the equivalence.
More precisely, the algebra of functions on the quotient is replaced by
a noncommutative algebra of functions on the graph of the equivalence
relation with the associative convolution product dictated by the
transitivity property of the equivalence relation,
$$ (f_1\star f_2) (x,y) = \sum_{x\sim z\sim y} f_1(x,z) f_2(z,y). $$

\smallskip

More precisely, in the case of the action of a discrete group $G$ on
a (compact) topological space $X$, the resulting algebra of (continuous) 
functions on the quotient is the {\em crossed product algebra}
$C(X) \rtimes_\alpha G$, where the associative, 
noncommutative product is given by
$(f U_g) (h U_{g'})= f \alpha_g(h) U_{gg'}$, with $\alpha_g(h)(x)=
h(g^{-1}(x))$.

\medskip

In the case of the quotient of $S^1$ by the action of $\Z$ generated
by $\exp(2\pi i \theta)$, an irrational rotation $\theta \in \R \smallsetminus \Q$,
the quotient is therefore described by the algebra $C(S^1)\rtimes_\theta \Z$.
This is by definition the algebra $\cA_\theta$ of continuous functions on the 
noncommutative torus $\bT_\theta$ of modulus $\theta$. 

\smallskip

An equivalent description of the irrational rotation
algebra $\cA_\theta$ is as the universal $C^*$-algebra generated 
by two unitaries $U$, $V$ with the commutation relation $VU =e^{2\pi i
\theta} UV$. It has a smooth structure given by the smooth subalgebra
of series $\sum_{n,m} a_{n,m} U^n V^m$ with rapidly decaying
coefficients (\cf \cite{CoCR}). 

\smallskip

Morita equivalence is the correct notion of
isomorphism for noncommutative spaces, and it can be formulated in
terms of the existence of a bimodule that implements an equivalence between
the categories of modules for the two algebras. The
algebras $\cA_{\theta_1}$ and $\cA_{\theta_2}$ are Morita equivalent
if and only if there exists a $g\in \SL_2(\Z)$ acting on $\R$ by
fractional linear transformations, such that $\theta_1=g\theta_2$, see
\cite{CoCR}, \cite{Rie}.  The bimodules realizing the
Morita equivalences between noncommutative tori are obtained
explicitly in \cite{CoCR} in terms of spaces of Schwartz functions on the
line, and in \cite{Rie} via a construction of projectors.

\medskip

One can also describe the irrational rotation algebra of the noncommutative
torus as a twisted group algebra $C^*(\Z^2,\sigma_\theta)$, with the
cocyle
\begin{equation}\label{cocycle}
 \sigma_\theta((n,m),(n',m'))= \exp(-2\pi i (\xi_1  nm' + \xi_2 m n')),
\end{equation} 
with $\theta=\xi_2-\xi_1$.
This is the norm closure of the action of the twisted group ring on $\ell^2(\Z^2)$
with the generators $U$ and $V$ are given by
$$ U f(n,m) = e^{-2\pi i \xi_2 n} f(n,m+1), \ \ \ \  V f(n,m) = e^{-2\pi i \xi_1 m} f(n+1,m). $$
This description of the noncommutative torus is especially useful in the
noncommutative geometry models of the integer quantum Hall effect, where
this noncommutative space replaces the Brillouin zone in the presence of a
magnetic field, see \cite{Bell}, \cite{MarMat}.

\section{$L$-functions, solvmanifolds, and noncommutative tori}

I give an overview here of recent progress in understanding the
geometry of a special class of noncommutative tori, which have
real multiplication, realized by nontrivial self Morita equivalences.
These are the quantum tori $\bT_\theta$ with $\theta\in \R$ a
quadratic irrationality. In particular,  I will focus on a results 
from \cite{Mar} that realizes certain $L$-functions associated to
real quadratic fields in terms of Riemannian and Loretzian geometry
on the noncommutative tori with real multiplication.

\subsection{Noncommutative tori with real multiplication}

The starting observation of Manin's ``Real multiplication program" is
the following. The elliptic curves with complex multiplication are the only
ones that have additional nontrivial endomorphisms, by the ring
of integers $O_\K$ of an imaginary quadratic field, and they correspond
to lattices $\Lambda \subset \C$ that are $O_\K$-submodules with 
$\Lambda \otimes_{O_\K} \K \cong \K$, which corresponds to the
parameter $\tau$ being a CM point of $\H$ for the imaginary quadratic field $\K$. 
In the same way, the noncommutative tori $\cA_\theta$ for which the
modulus $\theta$ is a real multiplication point in $\R$,  in
a real quadratic field $\K \subset \R$, have non-trivial self Morita 
equivalences, which play the role of the additional automorphisms of
the CM elliptic curve.

\subsection{Analytic versus algebraic model}\label{analalg} 

A good part of the recent work on noncommutative tori with
real multiplication was aimed at developing an algebraic
model for these objects, in addition to the analytic model
as quotients and crossed product algebras. 

\smallskip

The most interesting approach to algebraic models
for noncommutative tori is the one developed in
\cite{Poli}, which is based on enriching the bimodules
that give the self Morita equivalences with a 
``complex structure", in the sense of \cite{CoRie}. These are 
parameterized by the choice of an auxiliary elliptic curve $E$, or equivalently
by a modulus $\tau\in \H$ up to $\SL_2(\Z)$. By a
suitable construction of morphisms, one obtains in this way a category of holomorphic
vector bundles and a fully faithful functor to the derived category $D^b(E)$ 
of coherent sheaves on the auxiliary elliptic curve. The image is given 
by stable objects in the heart of a nonstandard t-structure, which depends 
on the parameter $\theta$ of the irrational rotation algebra $\cA_\theta$ 
of the noncommutative torus. The real multiplication gives rise to 
autoequivalences of $D^b(E)$ preserving the t-structure. 

\smallskip

This then makes it possible to associate to a noncommutative torus $\bT_\theta$
with real multiplication a noncommutative algebraic variety, in the sense
of \cite{AvB}. These are described by graded algebras of the form 
\begin{equation}\label{RMalg}
 A_{F,O}=\bigoplus_{n\geq 0} {\rm Hom} (O,F^n (O)) 
\end{equation}
where $O$ is an object of an additive category and $F$ is an additive functor. 
In the case of the noncommutative tori of \cite{Poli}, the additive category is the heart 
of the t-structure in $D^b(E)$, the object $O$ is $\cA_\theta$, and $F$ is 
induced by real multiplication, tensoring with the bimodule that 
generates the nontrivial self Morita equivalences. 

\smallskip

The resulting ring was then related in \cite{Vla} to the ring of quantum
theta functions. These provide a good theory of theta functions for 
noncommutative tori developed in \cite{Man-theta}, \cite{Man-theta2}. 
As in the case of the classical theta functions, these can be constructed
in terms of Heisenberg groups as a deformation of the classical case,
see \cite{Man-theta} (further elaborated upon in \cite{Plazas2}.)
The relation between the quantum theta functions and the
explicit construction of bimodules over noncommutative tori 
via projections was established in \cite{Boca}. 

\smallskip

The arithmetic properties of the algebras of \cite{Poli} were studied in \cite{Plazas},
in terms of an explicit presentation of the twisted homogeneous coordinate rings 
\eqref{RMalg} or real multiplication noncommutative tori, which involves 
modular forms of cusp type with level specified by an explicitly determined congruence subgroup.
A field of definition for these arithmetic structures on noncommutative tori can then
be specified in terms of the field of definition of the auxiliary elliptic curve. 
It is not yet clear whether this approach to defining algebraic models for
noncommutative tori with real multiplication can be successfully employed to
provide a substitute for the coordinates of torsion points of elliptic curves in the CM case.

\smallskip

There is, however, another approach which works directly with the
analytic model of noncommutative tori and with the candidate generators
for abelian extensions of real quadratic fields given by Stark numbers.

\subsection{Stark numbers and $L$-functions}

There is in number theory a conjectural candidate for
explicit generators of abelian extensions of real
quadratic fields, in the form of Stark numbers, \cite{Stark}.
These are obtained by considering a family of
$L$-functions associated to lattices $L \subset \K$ in
a real quadratic field.  In the notation of \cite{Man},
one considers an $\ell_0 \in O_\K$, with the property
that the ideals  $\fb=(L,\ell_0)$ and $\fa=(\ell_0)\fb^{-1}$ are
coprime with $\ff=L \fb^{-1}$. Let $U_L$ denote the set of
units of $\K$ such that $u (\ell_0 + L)= \ell_0 +L$, and let
$'$ denote the Galois conjugate, with $N(\ell)=\ell \ell'$.
One then considers the function
\begin{equation}\label{StarkZeta}
\zeta(L,\ell_0,s) = {\rm sign}(\ell_0') \,\, N(\fb)^s \, 
\sum_{\ell\in (\ell_0 +L)/U_L} \frac{{\rm sign}(\ell')}{|N(\ell)|^s} .
\end{equation}
The associated Stark number is then
\begin{equation}\label{Stark}
S_0 (L,\ell_0) = \exp ( \frac{d}{ds} \zeta(L,\ell_0,s)|_{s=0} ).
\end{equation}
Part of the ``real multiplication program" of \cite{Man}, \cite{Man2} is the
question of providing an interpretation of these numbers directly in terms 
of the geometry of noncommutative tori with real multiplication.

\medskip

To understand how one can relate these numbers to RM
noncommutative tori and to a suitable noncommutative space
that plays the role of the quotient $E/\Aut(E)$ of a CM elliptic curve,
we concentrate here on the case of a closely related $L$-function,
the Shimizu $L$-function of a lattice in a real quadratic field. 

\smallskip

The lattice $L\subset \K$ define a lattice $\Lambda=\iota(L) \subset\R^2$
via the two embeddings $L\ni \ell\mapsto (\ell,\ell')$. The
group $V$ of units of $\K$ satisfying 
$$ V= \{ u \in O_\K^* \,|\, u L \subset L, \ \iota(u)\in (\R^*_+)^2 \} $$
has generator a unit $\epsilon$ and it
acts on $\Lambda$ by $(x,y)\mapsto  (\epsilon x,\epsilon' y)$.
The Shimizu $L$-function is then given by
\begin{equation}\label{Shimizu}
L(\Lambda,s) = \sum_{\mu \in (\Lambda \smallsetminus \{ 0 \})/V} \frac{{\rm sign} (N(\mu))}{|N(\mu)|^s}.
\end{equation}
This corresponds to the case $\ell_0=0$ of \eqref{StarkZeta}, with the sum avoiding 
the point $0\in \Lambda$.

\subsection{Solvmanifolds and noncommutative spaces}

The hint on how the $L$-function \eqref{Shimizu} is related to RM noncommutative
tori comes from a well known result of Atiyah--Donnelly--Singer \cite{ADS}, which
proved a conjecture of Hirzebruch relating the Shimizu L-function to the signature
of the Hilbert modular surfaces, through the computation of the eta invariant of a
3-dimensional solvmanifold which is the link of an isolated singularity of the Hilbert 
modular surface. The result of \cite{ADS} is in fact more generally formulated for
Hilbert modular varieties and $L$-functions of totally real fields, but for our
purposes we concentrate on the real quadratic case only.  

\smallskip

Although it does not look like it at first sight, and it was certainly not formulated in those terms,
the result of \cite{ADS} is in fact saying something very useful about the geometry
of noncommutative tori with real multiplication, as I explained in \cite{Mar}.

\smallskip

A first observation is the fact that, in noncommutative geometry, one often has a 
way to construct a commutative model, up to homotopy, of a noncommutative space
describing a bad quotient. The idea is similar to the use of homotopy quotients
in topology, and is closely related to the Baum--Connes conjecture. In fact, the
latter can be seen as the statement that invariants of noncommutative spaces,
such as $K$-theory, can be computed geometrically using a commutative model
as homotopy quotient. 

\smallskip

As we recalled above, a ``bad quotient" can be described by 
a noncommutative space with algebra of functions an
associative convolution algebra, the crossed product algebra
in the case of a group action. In particular, we consider the
noncommutative space describing the quotient $\bT_\theta /\Aut(\bT_\theta)$
of a noncommutative torus with real multiplication by the automorphisms
coming from the group $V$ of units in the real quadratic field $\K$ 
preserving the lattice $L\subset \K$. The quotient of the action of the group 
of units $V$ on the noncommutative torus with real multiplication 
is described by the crossed product algebra $\cA_\theta \rtimes V$. This can 
also be described by a twisted group algebra of the form
\begin{equation}\label{twistsolvalg}
\cA_\theta \rtimes V \cong C^*(\Z^2 \rtimes_{\varphi_\epsilon} \Z, \tilde\sigma_\theta),
\end{equation}
where, after identifying the lattice $\Lambda$ with $\Z^2$ on a given basis, the
action of the generator $\epsilon$ of $V$ on $\Lambda$ is implemented by a matrix
$\varphi_\epsilon \in \SL_2(\Z)$, and one correspondingly identifies the
semidirect product $S(\Lambda,V)=\Lambda \rtimes_\epsilon V$ with  
$\Z^2 \rtimes_{\varphi_\epsilon}\Z$.  The cocycle $\tilde\sigma$ is given by
\begin{equation}\label{cocycle2}
\tilde\sigma_\theta ((n,m,k),(n',m',k'))= \sigma_\theta ((n,m),(n',m')\varphi_\epsilon^k).
\end{equation}
This is indeed a cocycle for $S(\Lambda,V)$, for $\xi_2 =-\xi_1=\theta/2$,
since in this case \eqref{cocycle} satisfies
$\sigma((n,m)\gamma, (n',m')\gamma)=\sigma((n,m),(n',m'))$, for
$\gamma \in \SL_2(\Z)$. 

\smallskip

Groups of the form $S(\Lambda,V)$ satisfy the Baum--Connes conjecture.
This implies that the quotient noncommutative space $\bT_\theta /\Aut(\bT_\theta)$,
with algebra of coordinates $C^*(S(\Lambda,V),\tilde\sigma_\theta)$, admits a
good homotopy quotient model. In this case, as shown in \cite{Mar}, this
homotopy quotient can be identified explicitly as the 3-dimensional smooth 
solvmanifold $X_\epsilon$ obtained as the
quotient of $\R^2 \rtimes_\epsilon \R$ by the group $S(\Lambda,V)$. This is the
same 3-manifold that gives the link of the singularity in the Hilbert modular surface
in \cite{ADS}, whose eta invariant computes the signature defect.

\smallskip

Another way to describe this 3-dimensional solvmanifold, with its natural metric, 
is in terms of Hecke lifts of geodesics to the space of lattices (see \cite{Man} and \cite{Mar}).
For $t\in \R$ one considers the lattice in $\R^2$ of the form
$$ \iota_t(L)= \{ (x e^t , y e^{-t}) \,|\, (x,y) \in \Lambda \}, $$
with $\iota_1(L)=\Lambda$, as above. Then one has a fibration
$T^2 \to S(\Lambda,V)\to S^1$, where the base $S^1$ is a circle of length $\log \epsilon$,
identified with the closed geodesic in $X_\Gamma(\C)$ corresponding to the
geodesic in $\H$ with endpoints $\theta,\theta'\in \R$, for $\{ 1, \theta \}$ a basis of
the real quadratic field $\K$. The fiber over $t\in S^1$ is the 2-torus 
$T^2_t=\R^2 /\iota_t(L)$.

\smallskip

The result of \cite{ADS} can then be reintepreted as saying that the spectral theory of
the Dirac operator on the 3-dimensional solvmanifold $X_\epsilon$ can be decomposed
into a contribution coming from the underlying noncommutative space $\bT_\theta /\Aut(\bT_\theta)$,
and an additional spurious part, which depends on the choice of a homotopy model for
this quotient. The part coming from the underlying noncommutative torus is the one
that recovers the Shimizu $L$-function and that is responsible for the signature defect
computed in \cite{ADS}. 

\subsection{Spectral triples}

To understand how the Dirac operator on the manifold $X_\epsilon$ can be 
related to a Dirac operator on the noncommutative space, one can resort to the
general formalism of {\em spectral triples} in noncommutative geometry \cite{CoS3}.
One encodes metric geometry on a noncommutative space by means of the
data $(\cA,\cH,D)$ of a representation on a Hilbert space $\cH$ 
of a dense subalgebra $\cA$ of the 
algebra of coordinates, together with a self-adjoint (unbounded) operator $D$ 
on $\cH$ with compact resolvent, satisfying the condition that commutators
$[D,a]$ with elements of the algebra are bounded operators. This plays the
role of an abstract Dirac operator which provides the metric structure.

\smallskip

\subsection{The Shimizu $L$-function and noncommutative tori}

One can then relate the Dirac operator on $X_\epsilon$ to a spectral triple
on the noncommutative torus with real multiplication, which recovers the
Shimizu $L$-function, in two steps, \cite{Mar}. The first makes use of the
isospectral deformations of manifolds introduced in \cite{CoLa}. Given
a smooth spin Riemannian manifold $X$, which admits an action of a torus
$T^2$ by isometries, one can construct a deformation of $X$ to a family
of noncommutative spaces $X_\eta$, parameterized by a real parameter 
$\eta\in \R$, with algebras of coordinates $\cA_{X_\eta}$, in such a way 
that, if $(C^\infty(X), L^2(X,S), D)$ is the
original spectral triple describing the ordinary spin geometry on $X$, then
the data $(\cA_{X_\eta}, L^2(X,S), D)$ still give a spectral triple on $X_\eta$.
In this way, one can isospectrally deform the fibration $T^2 \to X_\epsilon \to S^1$ to a
noncommutative space $X_{\epsilon,\theta}$, which is a fibration
$\bT_\theta \to X_{\epsilon,\theta} \to S^1$, where $\bT_\theta$ is the 
noncommutative torus with real multiplication. One then checks that, up
to a unitary equivalence, the restriction of the Dirac operator to the fiber
$\bT_\theta$ gives a spectral triple on this noncommutative torus with
Dirac operator of the form 
\begin{equation}\label{DiracTtheta}
D_{\theta,\theta'} =\left( \begin{array}{cc} 0 & \delta_{\theta'} -i \delta_\theta \\
\delta_{\theta'} + i \delta_\theta & 0 \end{array} \right),
\end{equation}
with $\{ 1, \theta \}$ the basis for the real quadratic field $\K$ and $\theta'$ the
Galois conjugate of $\theta$. The derivations $\delta_\theta$ and $\delta_{\theta'}$
act as
$$ \delta_\theta \psi_{n,m} = (n+m \theta) \psi_{n,m}, \ \ \  
\text{ and } \ \ \   \delta_\theta \psi_{n,m} = (n+m \theta') \psi_{n,m}, $$
and they correspond to leafwise derivations $e^t \partial_x$ and
$e^{-t} \partial_y$ on the tori $T^2_t$. The Dirac operator
$D_{\theta,\theta'}$ decomposes into a product of an operator with
spectrum ${\rm sign}(N(\mu)) |N(\mu)|^{1/2}$, which recovers the
Shimizu $L$-function, and a term whose spectrum only depends on
the powers $\epsilon^k$ on the unit $\epsilon$, see \S 7 of \cite{Mar}.

\smallskip

\subsection{Lorentzian geometry}

An important problem in the context of noncommutative geometry is
extending the formalism of spectral triples from Riemannian to Lorentzian
geometries. This is especially important in the particle physics and
cosmology models  based on spectral triples and the spectral action
functionals, see \cite{CCM}, \cite{MaPie}. A proposal for Lorentzian
noncommutative geometries, based on Krein spaces replacing Hilbert
spaces in the indefinite signature context, was developed in \cite{Stro}.

\smallskip

Another interesting aspect of the geometry of noncommutative tori with 
real multiplication is the fact that the spectral triples described above
admit a continuation to a Lorentzian geometry, based on considering
the norm of the real quadratic field $N(\lambda)=\lambda_1 \lambda_2=
(n+m\theta)(n+m \theta')$ as the analog of the wave operator in
momentum space, with modes $\Box_\lambda = N(\lambda)$. The
Krein involution is constructed using the Galois conjugation of the
real quadratic field, and the Wick rotation to Euclidean signature of
the resulting Lorentzian Dirac operator $\cD_{\bK}$ on $\bT_\theta$,
with $\cD_{\bK,\lambda}^2 = \Box_\lambda$, recovers the
Dirac operator $D_{\theta,\theta'}$. The eta function of the Lorentzian
spectral triple is a product
$$ \eta_{\cD_{\bK}}(s) = L(\Lambda,V,s) Z(\epsilon,s), $$
of the Shimizu $L$-function and a function that only depends on the
unit $\epsilon$.

\subsection{Quantum field theory and noncommutative tori}

This result of \cite{Mar} recalled above explains how certain 
number theoretic $L$-functions associated to real quadratic
fields, such as the Shimizu $L$-function or, more generally,
the zeta functions of \eqref{StarkZeta} arise from the noncommutative
geometry of noncommutative tori with real multiplication $\bT_\theta$
and their quotients $\bT_\theta/\Aut(\bT_\theta)$. 

\smallskip

One would then like to explain the meaning in terms of
noncommutative geometry of numbers of the form
$\exp( L^\prime(0) )$, where $L(s)$ is one of
these $L$-functions, since this is the class of numbers
that the Stark conjectures propose as conjectural
generators of abelian extensions. While there is at
present no completely satisfactory answer to this
second question, I describe here some work in 
progress in which I am trying to provide such interpretation
in terms of quantum field theory.

\smallskip

It should not come as a surprise that one would aim at realizing
numbers of arithmetic significance in terms of quantum
field theory. In fact, there is a broad range of results (see
\cite{Mar-feymot} for an overview) relating the evaluation
of Feynman integrals in quantum field theory to the
arithmetic geometry of motives. 

\smallskip

Here the point of connection is the zeta function
regularization method in quantum field theory. 
This expresses the functional integral that gives the partition function as
$$ \int e^{-\langle \phi, D \phi \rangle} \cD[\phi]  \sim (\det(D))^{-1/2}, $$
where the quantity $\det(D)$ here is obtained through
the zeta function regularization, using the
zeta function $\zeta_D(s)=\Tr(|D|^{-s})$ of the operator $D$ and
setting $\det(D)=\exp(-\zeta_D^\prime(0))$.

\smallskip

To adapt this to the setting described above of spectral triples
on a noncommutative torus with real multiplication, one can
use the fact that there is a well developed method \cite{GJKW} for 
doing quantum field theory on finite projective modules,
that is, for fields that are sections of ``bundles over noncommutative
spaces". This formalism was developed completely explicitly
in  \cite{GJKW} for the case of finite projective modules over 
noncommutative tori. In the case with real multiplication, one
has a preferred choice of a QFT, namely the one
associated to the bimodule that generates the non-trivial self 
Morita equivalences that give RM structure. A description of
the numbers \eqref{Stark} in terms of this quantum field theory
is work in progress \cite{MarQFT}.

\section{The noncommutative boundary of modular curves}

In the case of the imaginary quadratic fields, as we mentioned
above, the other approach to constructing abelian extensions
is by considering, instead of individual CM elliptic curves, 
the CM points on the moduli space of elliptic curves. 

\smallskip

In terms of noncommutative tori, one can similarly consider
a moduli space that parameterizes the equivalence classes
under Morita equivalence. This itself is described by 
a noncommutative space, which corresponds to the
quotient of $\P^1(\R)$ by the action of $\Gamma=\SL_2(\Z)$
by fractional linear transformations. As a noncommutative
space, this is described by the crossed product algebra
$C(\P^1(\R))\rtimes \Gamma$. This space parameterizes
degenerate lattices where $\tau\in \H$ becomes a
point $\theta\in \R$. One thinks of this space as the ``invisible
boundary of the modular curve $X_\Gamma(\C)$. It
complements the usual boundary $\P^1(\Q)/\Gamma$ 
(the cusp point corresponding to the degeneration of
the elliptic curve $E_\tau(\C)$ to the multiplicative group
$\bG_m(\C)$) with the irrational points $(\R\smallsetminus\Q)/\Gamma$,
treated as a noncommutative space. These irrational points account for the degenerations
to noncommutative tori, ``invisible" to the usual world of algebraic
geometry but nonetheless existing as noncommutative spaces.

\smallskip

In this approach, the main question becomes identifying what
remnants of modularity one can have on this ``invisible
boundary" and what replaces evaluating a modular form at
a CM point in this setting. 

\smallskip

\subsection{Modular shadow play}

A phenomenon similar to the ``holography principle" (also known as AdS/CFT correspondence)
of string theory relates the noncommutative geometry of
the invisible boundary of the modular curves to the algebraic geometry of the
classical ``bulk space" $X_\Gamma(\C)$ (see \cite{ManMar3}). For example, it was shown in
\cite{ManMar} that the $K$-theory of the crossed product algebra
$C(\P^1(\R))\rtimes \Gamma$ recovers Manin's modular complex \cite{Man-par}, 
which gives an explicit presentation of the homology of the modular 
curves $X_\Gamma(\C)$.

\medskip

A way of inducing on the noncommutative boundary $\P^1(\R)/\Gamma$
a class of functions corresponding to modular forms on the bulk space
$X_\Gamma(\C)$ was given in \cite{ManMar}, \cite{ManMar2} 
in terms of a L\'evy--Mellin transform, which can be thought of as
creating a ``holographic image" of a modular form on the boundary.

\smallskip

Consider a complex valued function $f$ which is defined on pairs $(q,q')$ of
coprime integers $q\geq q'\geq 1$, satisfying $f(q,q')=O(q^{-\epsilon})$ for 
some $\epsilon >0$. For $x\in (0,1]$ set
$$ \ell(f)(x)= \sum_{n=1}^\infty f(q_n(x), q_{n-1}(x)), $$
where the $q_n(x)$ are successive denominators of the continued
fraction expansion of $x$. L\'evy's lemma (see \cite{ManMar}) shows
that one has
$$ \int_0^1 \ell(f)(x) dx = \sum_{q\geq q'\geq 1; (q,q')=1} \frac{f(q,q')}{q(q+q')}. $$
This identity can be used to recast identities of modular forms in terms of
integrals on the boundary $\P^1(\R)$. For example, it is shown in \cite{ManMar}
that one can use the function  
$$ f(q,q')=\frac{q+q'}{q^{1+t}} \{ 0, q'/q\} $$
with $\Re(t)>0$ and $\{ 0, q'/q\}$ the classical modular symbol, 
together with the identity of \cite{Man-par},
$$ \sum_{d|m} \sum_{b=1}^d \int_{\{0,b/d\}} \omega = (\sigma(m)-c_m) \int_0^{i\infty} \pi^*_\Gamma(\omega), $$
where $\pi_\Gamma^*(\omega)/dz$ is a cusp form for $\Gamma=\Gamma_0(p)$,
with $p$ a prime, which is
an eigenvector of the Hecke operator $T_m$ with eigenvalue $c_m$,  
with $p\not| m$, and with $\sigma(m)$ the sum of the divisors of $m$. One then
obtains an identity of the form
$$ \int_0^1 dx \sum_{n=0}^\infty \frac{q_{n+1}(x)+ q_n(x)}{q_{n+1}(x)^{1+t}} 
\int_{\{ 0, q_n(x)/q_{n+1}(x)\} } \omega = 
\left(   \frac{\zeta(1+t)}{\zeta(2+t)} - \frac{L_\omega^{(p)}(2+t)}{\zeta^{(p)}(2+t)^2} \right) 
\int_0^{i\infty} \pi^*(\omega), $$
where $L_\omega^{(p)}$ and $\zeta^{(p)}$ are the Mellin transform and zeta
function with omitted $p$-th Euler factor. Other such examples were given in
\cite{ManMar}, \cite{Mar-lyap}.

\smallskip

This type of identities, recasting integrals of cusp forms on modular symbols in
terms of integrals along the invisible boundary of a transform of the modular form producing
a function on the boundary, can be formulated more generally and more abstractly 
as a way of obtaining ``shadows" of modular forms on the boundary. 
In \cite{ManMar2} one considers pseudomeasures associated to pair of rational points
on the boundary with values in an abelian group, $\mu: \P^1(\Q)\times \P^1(\Q) \to W$,
satisfying $\mu(x,x)=0$, $\mu(x,y)+\mu(y,x)=0$, and $\mu(x,y)+\mu(y,z)+\mu(z,x)=0$.
In particular the modular pseudomeasures satisfy
$\mu \gamma(x,y)=\gamma \mu(x,y)$, or an analogous
identity twisted by a character, 
where $\gamma(x,y)=(\gamma(x),\gamma(y))$ is the action of a finite index $\Gamma \subset
\SL_2(\Z)$ on $\P^1(\Q)$ by fractional linear transformations. 
The classical Hecke operators act on modular
pseudomeasures. 
Pseudomeasures can be equivalently formulated in terms of currents on the tree $\cT$ of $\PSL_2(\Z)$
embedded in the hyperbolic plane $\H$. In terms of noncommutative spaces, they can also 
be described as group homomorphisms $\mu: K_0(C(\partial\cT)\rtimes \Gamma)\to W$.

\smallskip

Integration along geodesics in $\H$ 
of holomorphic functions vanishing at cusps define pseudomeasures on the boundary. 
It is shown in \cite{ManMar2} that one can obtain ``shadows" of modular symbols on the
boundary by the following procedure. Given a finite index subgroup $\Gamma \subset \SL_2(\Z)$
and a weight $w\in \N$, let $\cS_{w+2}(\Gamma)$ be the $\C$-vector space of cusp forms $f(z)$
of weight $w+2$ for $\Gamma$, holomorphic on $\H$ and vanishing at cusps. 
Let $\cP_w$ be the space of homogeneous polynomials of degree $w$ in 
two variables and let $W$ be the space
of linear functionals on $\cS_{w+2}\otimes \cP_w$. Then 
$$ \mu(x,y) : f\otimes P \mapsto \int_x^y f(z) P(z,1) dz $$
defines a $W$-valued modular pseudomeasure, which is the shadow of the higher weight 
modular symbol of \cite{Sho}.

\smallskip

A general formulation is the given in \cite{ManMar2}, which 
encompasses the averaging techniques over successive convergents 
of the continued fraction expansion, used in \cite{ManMar} to relate Mellin transforms of weight-two 
cusp forms to quantities defined entirely on the noncommutative boundary of the modular curves. 
One considers a class of functions 
$\ell(f) (x)= \sum_I f(I) \chi_I(x)$ that are formal infinite linear combinations of characteristic
functions of ``primitive intervals" in $[0,1]$, with coefficients $f(I)$ in an abelian group.
More generally, this may depend on an additional regularization parameter, $\ell(f)(x,s)$.
The primitive intervals are those of the form $I=(g(\infty),g(0))$ with $g\in \GL_2(\Z)$. 
Pseudomeasures are completely determined by their values on these intervals.
The L\'evy--Mellin transform is then defined in \cite{ManMar2} as 
$\cL\cM(s) =\int_0^{1/2} \ell(f)(x,s) dx$. The integration over $[0,1/2]$ instead of $[0,1]$
keeps symmetry into account. When applied to a pseudomeasure obtained
as the shadow of a modular symbol, for an $\SL_2(\Z)$-cusp form this gives back
the usual Mellin transform. 

\smallskip

The formalism of pseudomeasures was also used in \cite{Man-wave} to
describe modular symbols for Maass wave forms, based on the work of
Lewis--Zagier \cite{LeZa}. In particular, Manin gives in \cite{Man-wave}
an interpretation of the L\'evy--Mellin transform of \cite{ManMar2} as an
analog at arithmetic infinity (at the archimedan prime) of the $p$-adic
Mellin--Mazur transform.

\subsection{Modular shadows and the Kronecker limit formula}

Modular pseudomeasures with values in a $\Gamma$-module $W$, with $\Gamma=\PSL_2(\Z)$,
give rise to 1-cocycles, by setting $\phi^\mu_x(\gamma)=\mu(\gamma x, x)$.
The cocycle condition $\phi(\gamma_1 \gamma_2)=\phi(\gamma_1)+\gamma_1 \phi(\gamma_2)$
follows from the modularity of $\mu$ together with the relations
$\mu(x,x)=0$, $\mu(x,y)+\mu(y,x)=0$, and $\mu(x,y)+\mu(y,z)+\mu(z,x)=0$, see \cite{ManMar2}.
Conversely, any cocycle with $\phi(\sigma)=\phi(\tau)$, where $\sigma$ and $\tau$ are
the generators of order two and three of $\Gamma = \Z/2\Z \star \Z/3\Z$. In fact, a pseudomeasure
is determined by the relations $(1+\sigma)\mu(0,\infty)=0$ and $(1+\tau+\tau^2)\mu(0,\infty)=0$,
while a 1-cocycle is determined by the relations $(1+\sigma)\phi(\sigma)=0$ and $(1+\tau+\tau^2)\phi(\tau)=0$. 

An interesting recent result \cite{VlaZa} gives a construction of a modular pseudomeasure involved
in a higher Kronecker limit formula for real quadratic fields. The pseudomeasure takes values in
$C(\P^1(\R))$ with the action of $\Gamma$ of weight $2k$. One considers a function 
$$ \psi_{2k}(x) ={\rm sign}(x) \sum_{p,q\geq 0}^* (p|x|+q)^k, $$
where the $*$ on the sum means that the sum is for $(p,q)\neq (0,0)$ and that the terms with
$p=0$ or $q=0$ are counted with a coefficient $1/2$. The modular pseudomeasure is given by 
setting $\mu(0,\infty)=\psi_{2k}$, since $\psi_{2k}=\phi(\sigma)=\phi(\tau)$ determines a 
1-cocycle. For $x>0$ the function $\psi_{2k}(x)$ is also expressed in terms of the derivatives
of the functions $\cF_{2k}$, constructed in terms of the digamma function $\Gamma'(x)/\Gamma(x)$,
which give the higher Kronecker limit formula proved in \cite{VlaZa} as
$$ \zeta(\fb,k) = \sum_{Q\in {\rm Red}(\fb)} (\cD_{k-1} \cF_{2k})(Q), $$
where $\zeta(\fb,s)=\sum_{\fn\in \fb} N(\fn)^{-s}$ and ${\rm Red}(\fb)$ is the 
set of reduced quadratic forms in the class $\fb$, by
seeing narrow ideal classes as $\Gamma$-orbits on the set of  
integer quadratic forms. The $\cD_{k-1}$ are differential operators of
order $k$ mapping differentiable functions of one variable to functions of
two variables, given explicitly in \cite{VlaZa}.

\smallskip

In particular, as shown in \cite{VlaZa}, one can use this higher Kronecker limit
formula to evaluate Stark numbers, as values at $k=1$ of the zeta-functions
rather than as derivatives at zero. This provides an alternative way of connecting
Stark numbers to the geometry of noncommutative tori, not by working with a
single noncommutative torus with real multiplication, but with their
noncommutative moduli space and the modular shadows.

\subsection{Quantum modular forms} There is at present another 
approach to extending modularity to the boundary, in a form that arises frequently
in very different contexts, such quantum invariants of 3-manifolds. Zagier recently
developed \cite{Za} a notion of {\em quantum modular forms}, which encompasses all these
phenomena. The idea is that, instead of the usual properties of modular forms,
namely a holomorphic function on $\H$ satisfying the modularity property
$$  (f|_k\gamma)(z):= f(\frac{az+b}{cz+d}) (cz+d)^{-k} = f(z), $$
one has a function $f$ defined on $\P^1(\Q)$, for which the function
\begin{equation}\label{hgamma}
 h_\gamma(x) = f(x) - (f|_k\gamma)(x), 
\end{equation}
which measures the failure of modularity, extends to a continuous or even
(piecewise) analytic function on $\P^1(\R)$.  

\smallskip

A more refined notion of ``strong" quantum modular form prescribes that,
besides having evaluations at all rational points, the function $f$ also has
a formal Taylor series expansion at all $x\in \Q$, and \eqref{hgamma} is
an identity of formal power series.
Typical examples of strong quantum modular forms described in \cite{Za}
have the additional property that the function $f: \P^1(\Q)\to \C$ extends
to a function $f: (\C\smallsetminus \R)\cup \Q \to \C$, which is analytic
on $\C\smallsetminus \R$, and whose asymptotic expansion approaching
a point $x\in \Q$ along vertical lines agrees with the formal Taylor series
of $f$ at $x$. Such quantum modular forms can be thought of as 
two analytic functions, on the upper and lower half plane, respectively,
that communicate across the rational points on the boundary. 

\smallskip

There are two observations one can make to relate this setting to
noncommutative geometry. One is that, in the case of quantum
modular forms, one is dealing with functions $f$ defined on the rational
points of the boundary, while the ``invisible boundary" consisting of the
irrational points is seen only through the associated function $h_\gamma$
which measures the failure of modularity of $f$. Thus, the object that
should be interpreted in terms of the noncommutative space
$C(\P^1(\R))\rtimes \Gamma$ is the $h_\gamma$ rather than the
quantum modular form $f$ itself. 

\smallskip

Another observation is that a similar setting, with functions that have
evaluations and Taylor expansions at all rational points, is provided 
by the Habiro ring of  ``analytic functions of roots of unity" \cite{Hab}.
This was, in fact, also developed to deal with the same phenomenology
of quantum invariants of 3-manifolds, such as the Witten--Reshetikhin--Turaev
invariants, which typically have a value at each root of unity as well as 
a formal Taylor expansion, the Ohtsuki series. Those strong
quantum modular forms that satisfy an additional integrality condition
needed in the construction of the Habiro ring
may be thought of as objects satisfying
a partial modularity property (through the associated $h_\gamma$)  
among these analytic functions of roots of unity. Several significant
examples of quantum modular forms 
given in \cite{Za} indeed define elements in the Habiro ring.

\smallskip

The functions in the Habiro ring were recently interpreted in \cite{ManF1}
as providing the right class of functions to do analytic geometry over the
``field with one element" $\F_1$. This was then reformulated in the
setting of noncommutative geometry in \cite{Mar-end} using the
notion of endomotives developed in \cite{CCM2} (see also \S 4 of \cite{CoMa-book})
which is a category of noncommutative spaces combining Artin motives
with semigroup actions, together with the relation between the endomotive
associated to abelian extensions of $\Q$ and Soul\'e's notion of
geometry over $\F_1$, established in \cite{CCM3}. The same
noncommutative space and some natural multivariable generlizations
are related in \cite{Mar-end} to another notion of geometry over
$\F_1$ developed by Borger \cite{Bor} in terms of consistent
lifts of Frobenius encoded in the structure of a $\Lambda$-ring.

\section{Quantum statistical mechanics and number fields}

The description of the boundary of modular curves in terms
of the noncommutative space $C(\P^1(\R))\rtimes \Gamma$,
for $\Gamma$ a finite index subgroup of the modular
group, accounts for degenerations of lattices with level
structures, to degenerate lattices (pseudolattices in the
terminology of \cite{Man}). In the adelic description, this
would correspond to degenerating lattices with level
structures at the archimedean component. In fact, one
can also consider degenerating the level structures at
the non-archimedean components. This leads to another
noncommutative space, which contains the usual
modular curves, and which also contains in its 
compactification the invisible boundary described above. 

\smallskip

In \cite{CoMa} such a noncommutative space of adelic
degenerations of lattices with level structures was
described as the moduli space of 2-dimensional
$\Q$-lattices up to commensurability and up to a
scaling relation. A $\Q$-lattice is a pair of a lattice
$\Lambda$ together with a group homomorphism
$\phi: \Q^2/\Z^2\to \Q\Lambda/\Lambda$ which
is a possibly degenerate level structure (it is not
required to be an isomorphism). Commensurability
means that $\Q\Lambda_1=\Q\Lambda_2$ and
$\phi_1=\phi_2$ modulo $\Lambda_1+\Lambda_2$.
The scaling is by an action of $\C^*$. The corresponding
noncommutative space is the convolution algebra
of functions $f((\Lambda,\phi),(\Lambda',\phi'))$ of 
pairs of commensurable lattices that are of degree
zero for the $\C^*$-action, with the convolution
product
$$ (f_1\star f_2)((\Lambda,\phi),(\Lambda',\phi')) =
\sum_{(\Lambda'',\phi'')\sim (\Lambda,\phi)} f_1((\Lambda,\phi),(\Lambda'',\phi''))
f_2((\Lambda'',\phi''),(\Lambda',\phi')). $$
This admits a convenient parameterization in terms
of coordinates $(g,\rho,z)$ with $g\in \GL_2^+(\Q)$,
$\rho\in M_2(\hat\Z)$, and $z\in \H$.

\smallskip 

The advantage of adopting this point of view is
that the resulting noncommutative space, whose algebra
of coordinates I denote here by $\cA_{GL(2),\Q}$, has a natural
time evolution, by the covolume of lattices
$$ \sigma_t(f) ((\Lambda,\phi),(\Lambda',\phi'))  = \left(\frac{{\rm covol}(\Lambda')}{{\rm covol}(\Lambda)}\right)^{it} \, f((\Lambda,\phi),(\Lambda',\phi')). $$

\smallskip

\subsection{Zero temperature states and modular forms}

The extremal low temperature KMS equilibrium states for
the dynamical system $(\cA_{GL(2),\Q},\sigma)$ are parameterized
by those $\Q$-lattices for which $\phi$ is an isomorphism (the invertible
ones). Thus the set of extremal low temperature KMS states can be
identified (\cite{CoMa}, \cite{CoMa-book} \S 3) 
with the usual Shimura variety $\GL_2(\Q) \backslash \GL_2(\A_\Q)/\C^*$.
This can be thought of as the set of the classical points of the noncommutative
space $\cA_{GL(2),\Q}$.

\smallskip

The adelic group $\Q^*\backslash \GL_2(\A_{\Q,f})$
acts as symmetries of this quantum statistical mechanical system, with the
subgroup $\GL_2(\hat\Z)$ of $\GL_2(\A_{\Q,f})=  \GL_2^+(\Q) \cdot \GL_2(\hat\Z)$
acting by automorphisms, and $\GL_2^+(\Q)$ by
endomorphisms, and the quotient by $\Q^*$ eliminating the inner symmetries
that act trivially on the KMS states.

\smallskip

The zero temperature extremal KMS states, defined in
\cite{CoMa} as weak limits of the positive temperature ones,
have the property that, when evaluated at elements of a
$\Q$-algebra $\cM_{GL(2),\Q}$ of unbounded multipliers
of $\cA_{GL(2),\Q}$, they give values that are evaluations
of modular forms $f\in F$ at points in $\H$. Under the identification
$\Q^*\backslash \GL_2(\A_{\Q,f})\cong \Aut(F)$, for
a generic set of points $\tau\in \H$ the action of
symmetries of the dynamical system is intertwined with
the action of automorphisms of the modular field. This is very
much like the $GL(1)$-case of \cite{BC}, which corresponds, 
in the same setting, to the case of 1-dimensional 
$\Q$-lattices.

\smallskip

\subsection{Imaginary quadratic fields}

One can recast in this setting of quantum statistical mechanical
systems the case of imaginary quadratic fields, \cite{CMR}. One
considers a similar convolution algebra for 1-dimensional $\K$-lattices,
for $\K=\Q(\sqrt{-d})$ and realizes it as a subalgebra $\cA_{\bK}$ of the algebra
of commensurability classes of 2-dimensional $\Q$-lattices recalled
above. In this case, the extremal low temperature KMS states are
parameterized by the invertible $\K$-lattices, which are
labelled by a CM point in $\H$ and an element in $\hat O_\bK$. The
evaluation of extremal zero temperature KMS states on the restriction
of the algebra $\cM_{GL(2),\Q}$ to $\cA_{\bK}$ then give evaluations
of modular forms at CM points and the action of symmetries induces
the correct action of $\Gal(\K^{ab}/\K)$.

\subsection{Quantum statistical mechanical systems for number fields}

The construction of \cite{CMR} of quantum statistical mechanical systems
$(\cA_\K, \sigma)$ associated to imaginary quadratic fields,
using the system for 2-dimensional $\Q$-lattices of \cite{CoMa}, was
generalized in \cite{HaPau} to a construction of a similar system
for an arbitrary number field, using a generalization of the $GL(2)$-system
to quantum statistical mechanical systems associated to arbitrary Shimura
varieties. Rewritten in the notation of \cite{LLN} these quantum
statistical mechanical systems $(\cA_\K,\sigma)$ for number fields are given by
semigroup crossed product algebras of the form 
$$  
\cA_\K = C(G^{ab}_{\K}\times_{\hat O_{\K}^*} \hat O_{\K})\rtimes J^+_{\K},
$$  
where $J^+_{\K}$ is the semigroup of integral ideals and $G^{ab}_{\K}=\Gal(\K^{ab}/\K)$.
These also admit an interpretation as convolution algebras of
commensurability classes of 1-dimensional $\K$-lattices, see
\cite{LLN}. The time evolution is by the norm of ideals 
$$  
\sigma_t (f) =f, \ \ \ \forall f \in C(G^{ab}_{\K}\times_{\hat O_{\K}^*} \hat O_{\K}),
\ \ \ \text{ and } \ \ \  \sigma_t (\mu_{\fn}) = N(\fn)^{it}\, \mu_{\fn}, \ \ \ \forall \fn \in  J^+_{\K}.
$$  

An explicit presentation for the algebras $\cA_\K$ was obtained in
\cite{CuLi}, by embedding them into larger crossed product algebras.
What is still missing in this general construction is the ``algebra of arithmetic
elements" replacing $\cM_{GL(2),\Q}$, on which to evaluate the zero temperature
extremal KMS states to get candidate generators of abelian extensions. In the
particular case of the real quadratic fields, such an algebra would contain the 
correct replacement for the modular functions on the invisible boundary of the
modular curves.

\smallskip

\subsection{Noncommutative geometry and anabelian geometry}

The quantum statistical mechanical systems for number fields described
above are explicitly designed to carry information on the abelian extensions
of the field, hence they involve the abelianization of the absolute Galois group.
However, it appears that these noncommutative spaces may in fact contain 
also the full ``anabelian" geometry of number fields. This is presently being
investigated in my joint work with Cornelissen \cite{CornMa}. The question
is to what extent one can reconstruct the number field from the system 
$(\cA_\K,\sigma)$. The fact that the partition function of this 
quantum statistical mechanical system is the Dedekind zeta function
and that the evaluation of low temperature KMS states on elements in
the algebra can be written in terms of Dirichlet series, shows that
at least the system recovers the arithmetic equivalence class of the
field. A similar results should in fact hold for function fields, where
a version of these quantum statistical mechanical systems in the
positive characteristic setting with partition function the Goss zeta
function was developed in \cite{ConsMa} (see \cite{Corn} for
the role of the Goss zeta function for arithmetic equivalence.)
It is more subtle to see whether the system $(\cA_\K,\sigma)$
recovers not only the field up to arithmetic equivalence but also
up to isomorphism, \cite{CornMa}.

\end{document}